\documentclass{RISEarticle} % -*- coding: utf-8 -*-
\RISEApprovedBy{Approved} % Default is "Preliminary" % Changing this for arXiv upload
\RISEPreparedBy{Pontus Johannisson} % Default is the author name, see below
\RISEDocumentIdentity{HAV-KK-MP} % Project/ID/reference...
\RISERevision{1}
\RISEDate{2022-11-07} % Empty for no date, commented out for today's date
\usepackage[frozencache]{minted}

\newcommand{\vsi}{VSI}
\newcommand{\vsifull}{vibration shock index}
\newcommand{\vsl}{VSL}
\newcommand{\vslfull}{vibration shock level}

\begin{document}

% The title and the full list of authors follows. (RISEPreparedBy is intended to
% be the "communicating" author, i.e., the person that has been responsible for
% the preparation of the document.)
\title{Definition and Quantification of Shock/Impact/Transient Vibrations}
\author{Pontus Johannisson, Hans Lindell\\[2mm]\small{RISE Research Institutes of Sweden}}
\date{} % Empty for no date, commented out for today's date
\maketitle

% The abstract is optional for short documents
\begin{abstract}
  Vibration injury in the hand-arm system from hand-held machines is one of the
  most common occupational health injuries and causes severe and often chronic
  nerve and vascular injury to the operator. Machines emitting shock vibrations,
  e.g., impact wrenches have since long been identified as a special risk
  factor. In legislative and standard texts the terms shock, impact, and
  transient vibration are frequently used to underline the special risk
  associated with these kinds of vibrations. In spite of this, there is no
  mathematically stringent definition what a shock vibration is or how the
  amplitude of the shock is defined.

  This lack of definitions is the subject of this article. This document
  discusses a number of candidate definitions for a \emph{\vsifull} (\vsi) that
  quantifies different vibration signals in terms of how localized they are in
  the time domain. The \vsi\ is intended to be used to classify and compare
  different vibration sources.

  The \vsi\ is independent of the vibration level, i.e., it is unchanged if the
  vibration signal is rescaled. The traditional root mean square method to
  determine the vibration level will not produce a value representative for the
  shocks occurring in a signal with high \vsi. Thus, there is a need for a
  complementing quantification method for the localized signal parts. Possible
  definitions for such a \emph{\vslfull} (\vsl) are suggested.

  A problem formulation is first stated together with a description of the
  approach used for designing the \vsi\ and the \vsl. After this, model signals
  are defined, which are used to discuss and evaluate the different candidate
  definitions. Then, a number of candidate definitions are discussed, leading up
  to a conclusion on which candidate definitions that are promising for
  experimental evaluation.
\end{abstract}

%% % The figure on the first page is optional and can potentially be replaced by an abstract an/or the table of contents.
%% \vfill % Vertical ``rubber space''
%% \begin{figure}[htbp]
%%   \begin{center}
%%     \includegraphics[height=50mm]{RISElogo.eps}\\
%%     \textit{Optional figure on the first page.}
%% %     \caption{No caption on the first page}
%% %     \label{fig:RISElogo}
%%   \end{center}
%% \end{figure}
%% \vfill % Vertical ``rubber space''

%% % The page breaks should be adjusted to the document at hand.
\newpage

\tableofcontents

\newpage

%%%%%%%%%%%%%%%%%%%%%%%%%%%%%%%%%%%%%%%%%%%%%%%%%%%%%%%%%%%%%%%%%%%%%% 
% Start of main text
\section{Introduction}
\label{introduction}
Humans are sensitive to mechanical vibrations, but the health effects depend on
many parameters, including the vibration amplitude/frequency and the exposure
time. One further aspect, which traditionally has not been given sufficient
attention, is the fact that some vibrating equipment generates potentially
harmful vibrations outside the frequency interval considered by the ISO~5349-1
standard~\cite{dandanell1986vibration, barregard2003hand, raju2011vibration}. These may lead to, e.g., nerve and blood vessel
injuries in fingers. In analogy with the term \emph{ultrasound}, these
vibrations have been named \emph{ultravibrations} and is a central concept of
this report.

Ultravibrations can be generated either by tools working at high frequency such
as, e.g., dental drills, or by tools that generate short pulses. One example of
the latter is impact wrenches, which generate shocks with a wide frequency
spectrum and periodically delivers power during short time intervals~\cite{lindell1998transient}. It is
clear that for extremely powerful pulses, even a single pulse can have a
quantifiable effect on human health. Thus, the total health effect could be
estimated from this health effect per pulse multiplied by the number of pulses
for which the user is exposed. If, on the other hand, a simple time average is
used, significant underestimation of the deleterious health effects may result
as this would fail to take the localized nature of the vibration into account.
The question is then when a vibration signal should be considered to be pulsed
or not.

This document has two main goals. The first is to suggest a method that answers
to what extent a given vibration signal should be considered to contain shocks.
This is done by introducing and discussing a number of candidate methods that
quantify how strongly localized a given acceleration signal is in the time
domain. This parameter is here referred to as the \emph{\vsifull} (\vsi) of the
vibration signal. The second goal is to quantify the vibration level in the
shocks of the signal, which will act as a complement to the traditionally used
root mean square (RMS) value. This is referred to as the \emph{\vslfull} (\vsl).

It is pointed out that these two goals are related but distinct. The \vsi\
answers the question to what extent the signal contains localized features and
this is a dimensionless quantity. The \vsl\ assigns a representative vibration
level value to the localized features. The dimensions of this quantity is the
same as for the signal. It should also be noted that quantifying the health
effects resulting from a certain vibration source is a separate topic, which is
outside the scope of this report. The \vsi\ and \vsl\ do not say anything direct
about health effects, but is of potential use as parameters in a method for
quantifying this.

The organization of this document is as follows. First, the problem is given a
clear formulation and a simplified physical model is used to introduce the
physical quantities the \vsi\ and \vsl\ will be based on. Then, model signals
are defined, which are used to evaluate the candidate \vsi\ and \vsl\ algorithms.
After this, a number of candidate definitions are given, together with a
discussion of their relative merits. The work is then concluded.

\section{\vsi\ and \vsl\ Definition Process}
The aim of this section is to give all background information needed to discuss
possible \vsi/\vsl\ candidate definitions. First, a number of required
properties for the \vsi/\vsl\ are specified, with the intention that these
properties will act as guiding principles during the development process. A
simple physical model is then introduced together with a problem formulation.
Simple model signals that are used to evaluate the candiate definitions are then
introduced. It is of central importance to later test the candidate definitions
using a large number of experimentally obtained vibration signals from different
tools in order to decide to what extent the \vsi/\vsl\ are valuable concepts in
practice, but this is outside the scope of the current document.

\subsection{\vsi\ and \vsl\ Selection Criteria}
\label{sec:Selection Criteria}
A \vsi/\vsl\ definition that is useful in practice is expected to have a number
of properties, which are listed in the following.
\begin{itemize}
\item Dimensions: The \vsi\ quantifies a property that is dimensionless. The
  \vsl\ has the same dimensions as the vibration signal.
\item Intuitive: The definitions should give results that are intuitively
  understandable, implying that, e.g., the following is true.
  \begin{itemize}
  \item If a vibration signal consists of a train of clearly visible pulses,
    then the \vsi\ should be high. If this signal is compared with one
    consisting of identical pulses with larger time separation, then the latter
    should have an even higher \vsi.
  \item The \vsl\ should be a value between zero and the maximum amplitude of
    the vibration signal. When there are shocks in the signal, the \vsl\ should
    be a value representative for the shocks.
  \end{itemize}
\item Relevant: The \vsi/\vsl\ should be based only on physical quantities that
  are clearly relevant for the health effects related to vibrating equipment. In
  order to be compatible with standardized vibration measurements, it can only
  take the acceleration signal as input.
\item Unambiguous: The definitions should be completely clear and to remove all
  subjectivity in the calculation process, the definitions must be given in
  terms of an algorithm that take the measured acceleration as input and produces
  the \vsi/\vsl\ as output.
\item Robust: The \vsi/\vsl\ should show robust numerical/statistical
  properties. This includes but is not limited to the following.
  \begin{itemize}
  \item The \vsi/\vsl\ should show a small sensitivity to noise and other
    imperfections, such as outliers, in the measured data.
  \item The \vsi/\vsl\ should converge in the sense that using a longer signal
    sequence should lead to more accurately estimated values.
  \item The difference in the \vsi/\vsl\ as estimated from two different
    measurements on the same vibration source should be small, given that both
    measurements are of good quality and sufficiently long.
  \end{itemize}
\item Parsimonious:\footnote{\url{https://en.wikipedia.org/wiki/Occam's_razor}}
  The definitions should contain as few arbitrary parameters and use as few
  assumptions as possible and be applicable to the widest possible range of
  experimentally obtained acceleration signals. When there are multiple
  suggestions with similar performance, preference should be given to the
  simplest approach.
\end{itemize}
It is also pointed out that the \vsi\ is a classification of the shape of the
waveform and should therefore not change if time or acceleration is rescaled.

\subsection{Physical Model}
\label{physical-model}
It is assumed that the vibrating system (including the human operator and the
material being machined) can be modeled as a driven harmonic oscillator. This is
very simplified since, e.g., (i) the real system is distributed and not
necessarily possible to describe using a simple lumped system, (ii) the
coefficients are here modeled to be independent of frequency, and (iii) there
may be additional terms, including nonlinear terms, in a model equation
describing the real system. Nevertheless, since it is not possible to know or
account for the differences between systems, this approach seems to be a
reasonable compromise. As is shown in Appendix~\ref{sec:Dissipated Energy from a
  Harmonic Oscillator}, the dissipated power is then proportional to the square
of the velocity of the vibration. Since the dissipated energy can be obtained by
integrating the power with respect to time, it may seem natural to base the
\vsi\ on this signal. However, there are a number of arguments against this
approach.
\begin{enumerate}
\item Although the velocity can, in principle, be obtained by integrating the
  measured acceleration, this may not be straightforward in practice for a
  number of reasons.
  \begin{itemize}
  \item The accelerometer noise will be integrated into a random walk and the
    variance for such a process increases linearly with time.
  \item In addition to noise, the accelerometer signal may be affected by bias
    drifts, nonlinearity etc. Integrating these types of errors may lead
    to a diverging velocity, i.e., a velocity error that grows with time.
  \item In addition to the tool acceleration, the accelerometer also measures
    gravity, which may be integrated into a diverging velocity.
  \end{itemize}
  Even though these problems can be mitigated by high-pass filtering of the
  acceleration, there is no guarantee that it is robust to base the \vsi\ on the
  velocity.
\item A more fundamental problem is that is not clear that the dissipated energy
  is the physical quantity that correlates best with the health effects from
  vibrations. It is, e.g., known that heating is not what causes damage to
  biological tissue from vibrations. It may be that the force is more relevant,
  in which case the \vsi\ should be based on the acceleration, and some results
  even indicate that it may be the time derivative of the force that is
  important.
\end{enumerate}

\subsection{Problem Formulation}
\label{sec:Problem Formulation}
If it would be assumed that (i) the described physical model is relevant and
(ii) the dissipated energy is the central concept, the previous loose discussion
can be replaced with a clear and intuitive criterion for a high \vsi: In a
vibration signal that is ``localized'' it is possible to find short time
intervals that contribute a large fraction of the total dissipated energy.

It is unfortunate that it is not known which is the most relevant physical
quantity to base the \vsi\ on. We choose to select the ``power
signal''\footnote{Citation marks are used here to indicate that there is no
  direct physical correspondence to the dissipated power. These citation marks
  will be dropped later in this document.} to be equal to the square of the
acceleration instead of the square of the velocity. It should be noted that it
is the square root of the mean value of this quantity that is the primary
quantity of standardized measurements~\cite[Section~4.4]{iso5349-1}, implying that
this choice brings the approach here closer to the existing standard.

With this modified definition of the power signal, the remainder of this report
will be based on the criterion above. Obviously, this must be made quantitative
in order to specify an algorithm, but this problem formulation will serve as a
guiding principle.

\subsection{Model Signals}
\label{model-signals}
Two model signals are used to produce example results from the candidate
\vsi/\vsl\ definitions. In practice, it will always be the case that the
vibrating equipment is stationary over long times, i.e., when the acceleration
is integrated two times, then the obtained position should oscillate around an
approximately fixed point. The easiest way to ensure this is to specify the
position and obtain the acceleration by a double differentiation with respect to
time and this is the method used here to define one ``continuous signal'' and
one ``pulsed signal''. In addition to these, white Gaussian noise (WGN) with RMS
value of one will be used as a model signal. A summary of the selected model
signals is given here and the calculations are found in Appendix~\ref{sec:Model
  Signal Calculations}.

\begin{figure}[p]
  \begin{center}
    \includegraphics[width=0.75\linewidth]{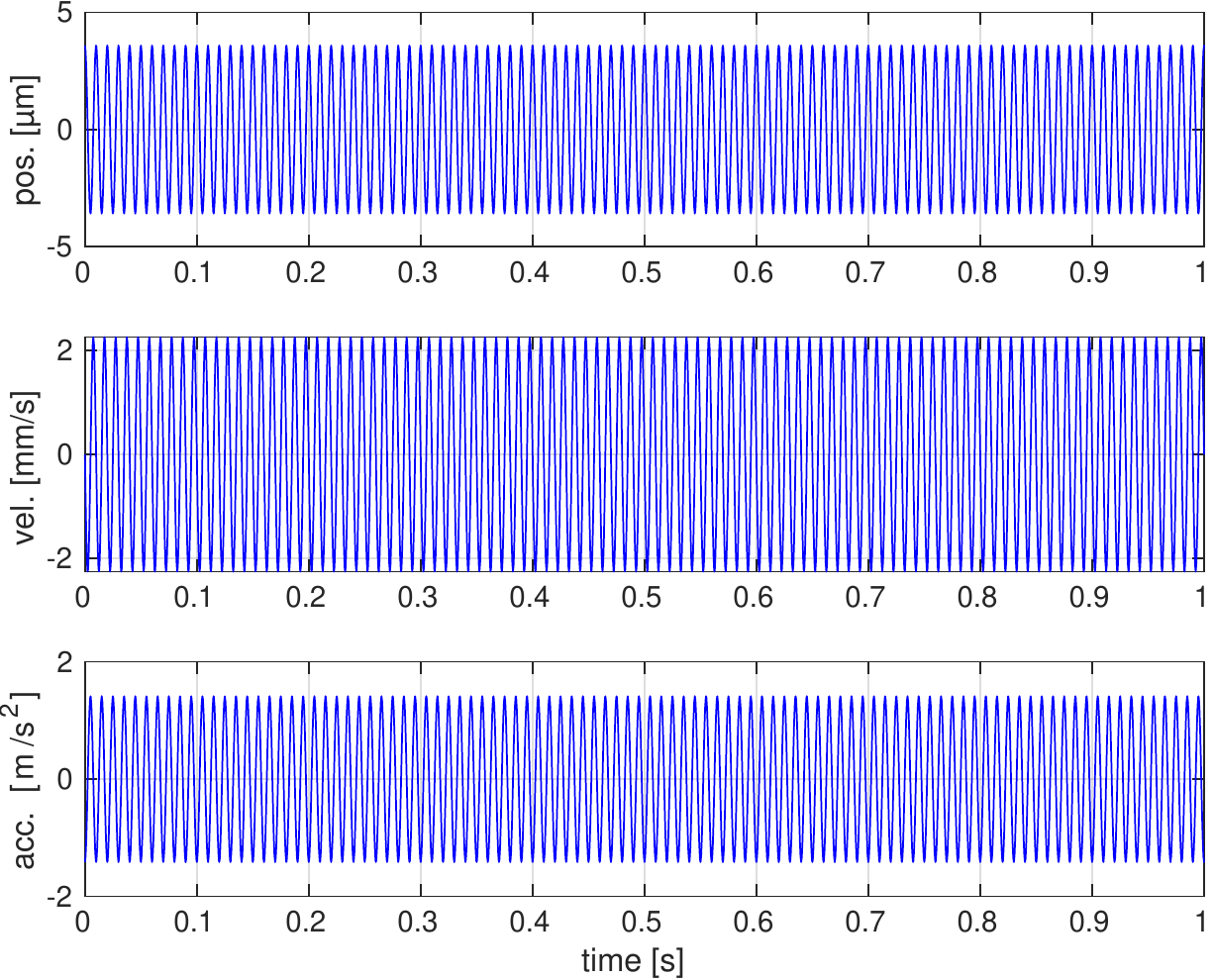}
    \caption{The continuous model signal.}
    \label{fig_continuous_signal}
  \end{center}
\end{figure}

As a model for a continuous signal, a harmonic oscillation is used according to
\begin{align}
  x_c(t) = A_c \cos(2 \pi f_c t + \phi_c),
\end{align}
where $A_c$ is the amplitude, $f_c$ is the frequency, and $\phi_c$ is the phase,
which is set to zero. As a convenient example value, choosing $A_c =
\SI{3.58}{\micro\meter}$ and $f_c = \SI{100}{Hz}$, the RMS value is one and the
signals in Fig.~\ref{fig_continuous_signal} are obtained. It is seen that the
peak acceleration $\hat{a}_c = A_c (2 \pi f_c)^2 \approx \SI{1.4}{m/s^2}$.

\begin{figure}[p]
  \begin{center}
    \includegraphics[width=0.75\linewidth]{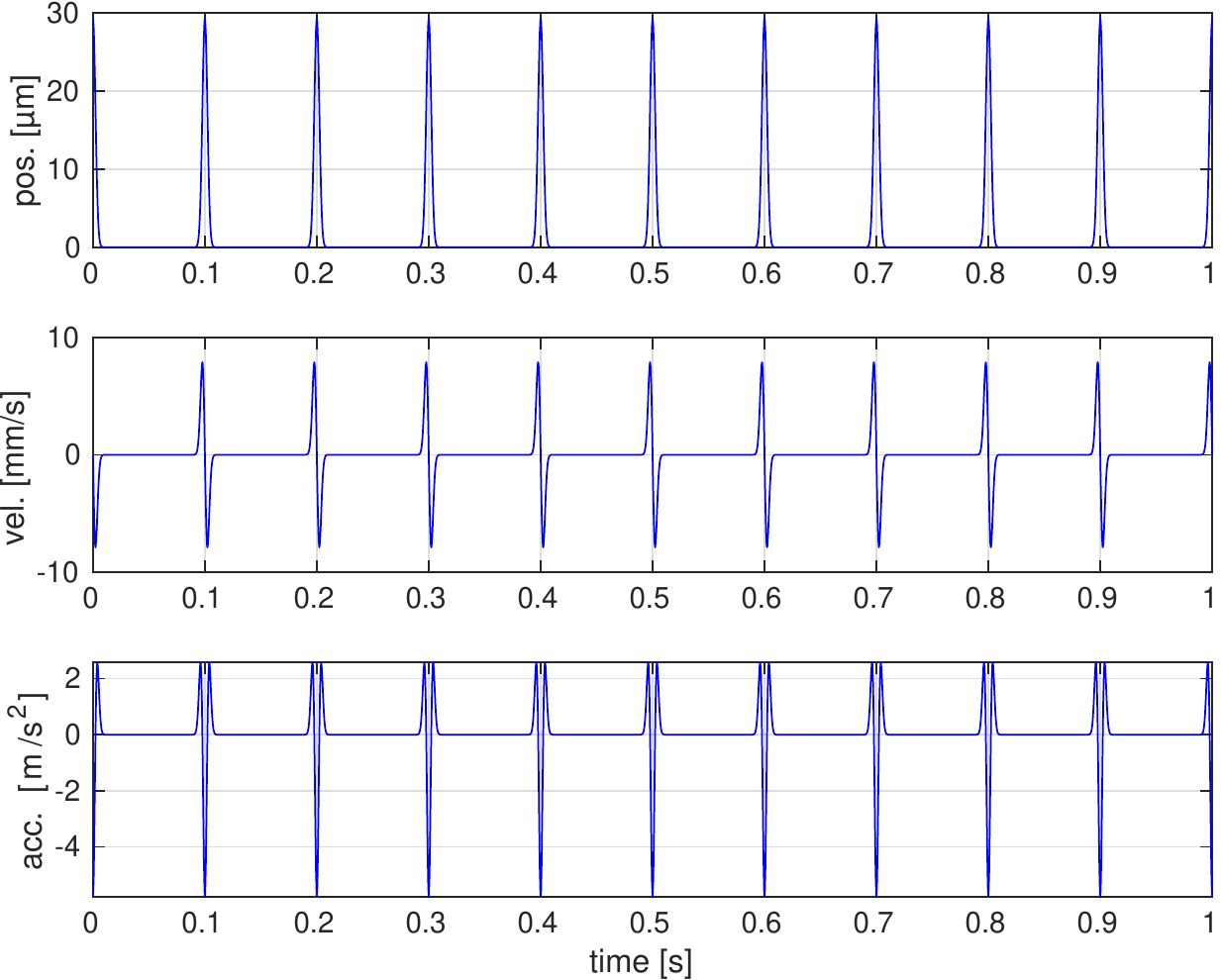}
    \caption{The pulsed model signal.}
    \label{fig_pulsed_signal}
  \end{center}
\end{figure}

As a model for a pulsed signal, a train of Gaussian pulses is used according to
\begin{align}
  x_p(t) = \sum_{k=-\infty}^{\infty} A_p \exp \left[ - \frac{(t - k T_p)^2}{2 t_p^2} \right],
\end{align}
where $A_p$ is the amplitude, $t_p$ is the pulse width, and $T_p$ is the time
separation between pulses. The pulse width $t_p$ is selected such that the
Fourier transform of $a_p$ has its maximum at $f_c$. The value for $T_p$ is left
as a free parameter, but for a signal with a high \vsi\ the pulse width should be
much smaller than the pulse separation, i.e., $t_p \ll T_p$. The value for $A_p$
is selected such that the root mean square (RMS) values are equal for the two
acceleration signals.\footnote{Note that a rescaling of the signal amplitude
  should not affect the \vsi\ and this condition is introduced simply in order to
  define the amplitude.} Setting $T_p = \SI{0.1}{s}$, the signals in
Fig.~\ref{fig_pulsed_signal} are obtained. In this case, the peak acceleration,
$\hat{a}_p = A_p/t_p^2 \approx \SI{5.8}{m/s^2}$, i.e., about four times the peak
value of the continuous signal.

\section{\vsi\ and \vsl\ Candidate Definitions}
In the following, a number of candidate definitions for how the \vsi\ and \vsl\
could be selected are given. Initially, the focus is on the \vsi. All these
suggestions are not promising, but the context provided by this description is
useful for discussing the relative merits of the different candidate
definitions.

\subsection{Using Pulse Identification}
From Figs.~\ref{fig_continuous_signal} and \ref{fig_pulsed_signal} it is obvious
that there are large differences in the visual appearance of the model signals.
This would suggest that, e.g., identifying the pulses and using their duty cycle
could be a useful approach. However, this parameter, which should be
proportional to $t_p/T_p$, would require unambiguous identification of the
pulses and their widths. While this is easy to do for the pulsed model signal,
it becomes significantly harder or even impossible if, e.g., there is a
considerable amount of noise, if the pulses are close to each other or have
multiple peaks, if the pulses have varying widths, and/or the pulse train is not
periodic. Thus, it seems difficult to find an algorithm with general validity that
identifies the pulses in the time domain.

In the frequency domain, the continuous model signal has a very narrow spectrum
and it may seem straightforward to define the \vsi\ in terms of the wide spectra
that arise for narrow time domain events like shocks. However, spectra for
periodic trains of pulses often have a complicated form, giving rise to similar
difficulties as using the pulse shapes in the time domain. Furthermore, it
should be noted that there are signals that have extremely broad spectra without
showing any pulsed behavior in the time domain. One example is WGN, which has a
flat spectrum that stretches over all frequencies.

From this, it is expected that while it is possible to describe some signals
directly from the shape of the time and/or frequency domain signals, it seems
difficult to formulate a general \vsi\ definition in this way.

\subsection{Using the Excess Kurtosis}
\label{sec:Using the Excess Kurtosis}
\begin{figure}[p]
  \begin{center}
    \includegraphics[width=0.7\linewidth]{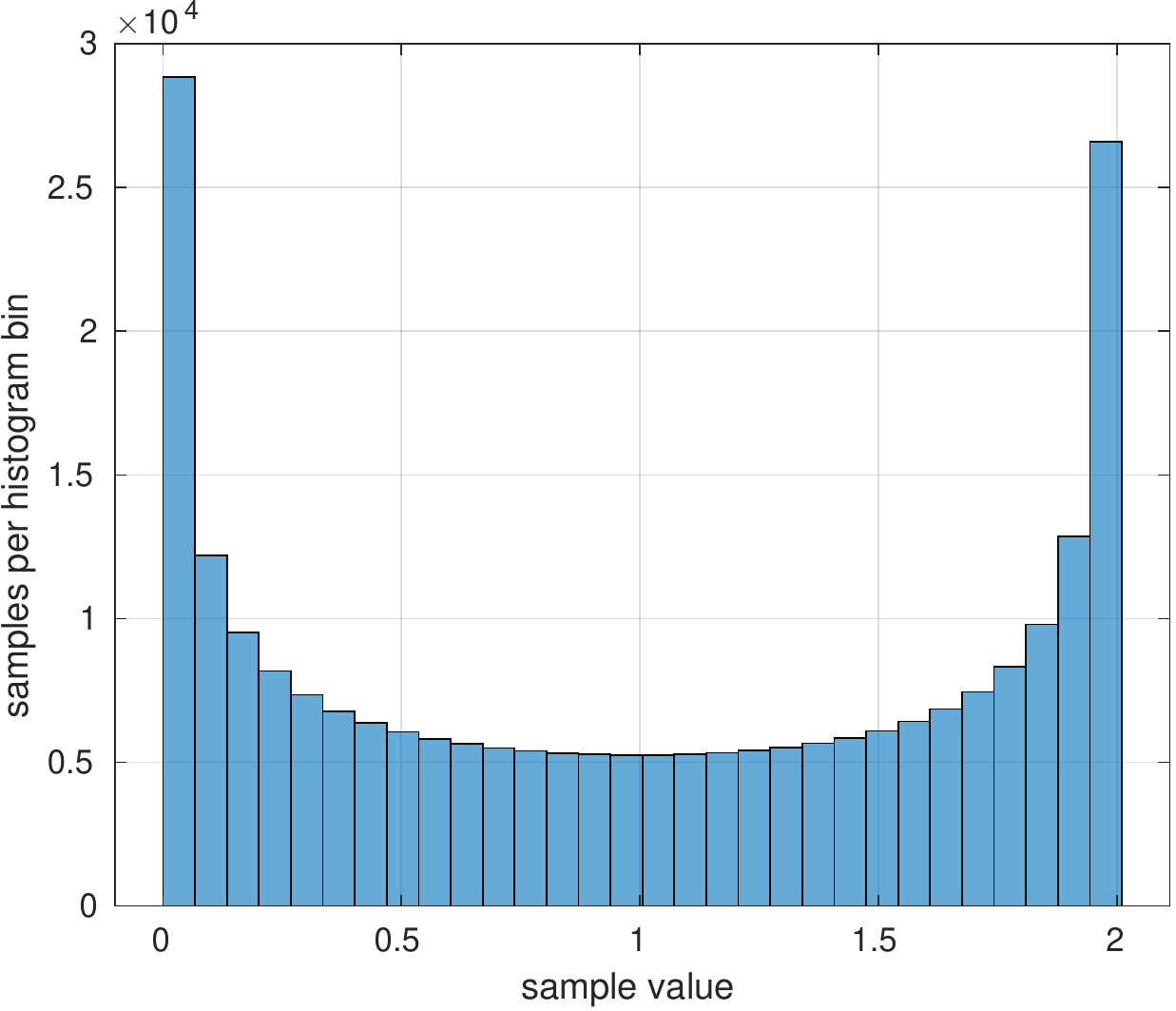}
    \caption{The histogram for the continuous power signal.}
    \label{fig_hist_continuous_signal}
  \end{center}
\end{figure}

\begin{figure}[p]
  \begin{center}
    \includegraphics[width=0.7\linewidth]{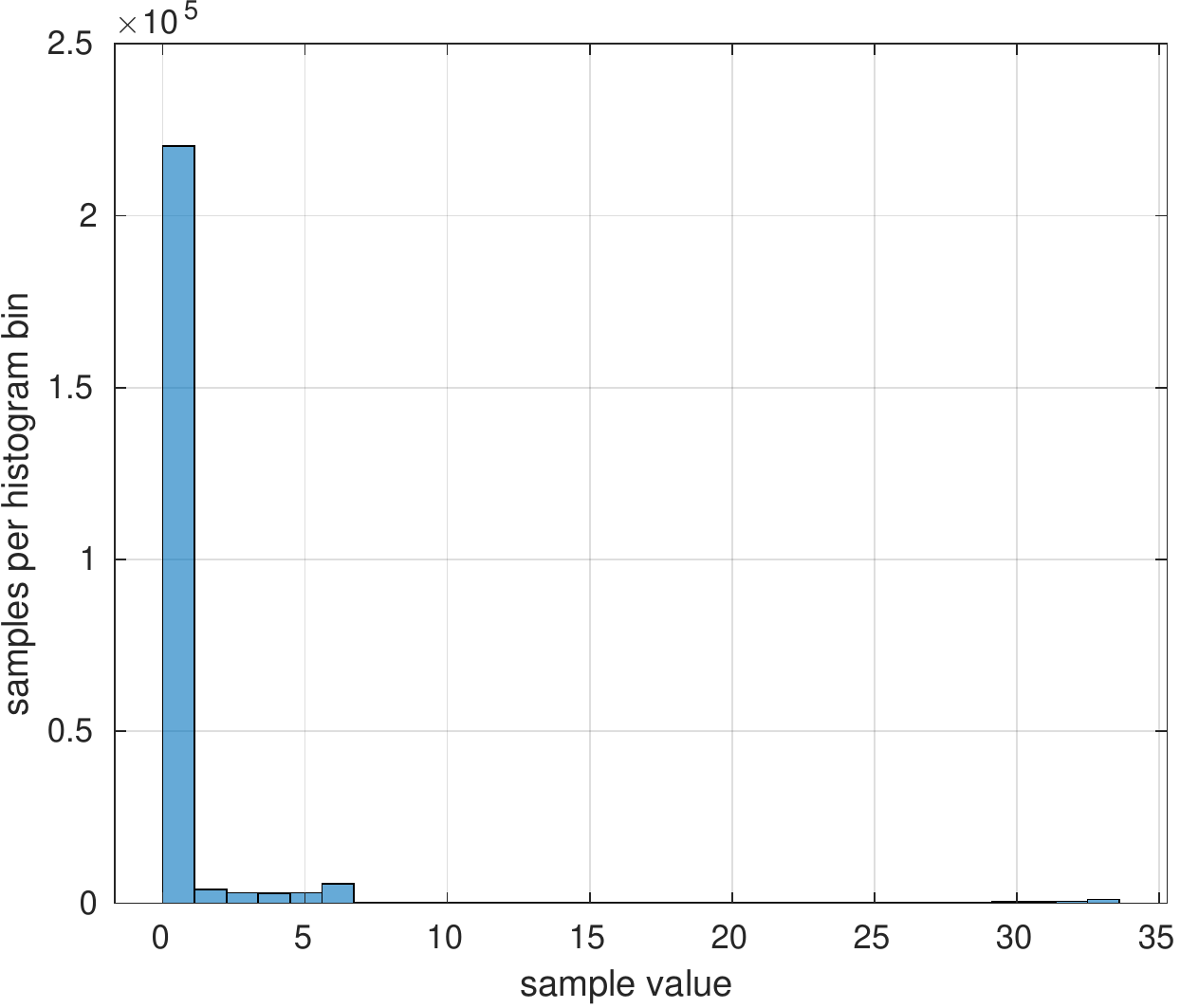}
    \caption{The histogram for the pulsed power signal. Note the barely visible
      samples with sample values above 30.}
    \label{fig_hist_pulsed_signal}
  \end{center}
\end{figure}

One idea is to use a statistical approach to quantify the distributions
of samples in the power signals. In order to discuss this, histograms for the
two model power signals, $a_c^2$ and $a_p^2$, are found in
Figs.~\ref{fig_hist_continuous_signal} and \ref{fig_hist_pulsed_signal}.
Although these may look counterintuitive, the properties of the figures are
straightforward to understand from $a_c(t)$ and $a_p(t)$.
\begin{itemize}
\item In the case of $a_c$, squaring the harmonic signal doubles the frequency
  and the oscillation occurs between zero and the maximum value. The most likely
  signal values are at the extreme points simply due to the fact that the signal
  is ``moving quickly'' through all other values. At the turning points,
  however, the signal has zero slope and this leads to a large number of samples
  close to these values.
\item In the case of $a_p$, on the other hand, the histogram shows that it is
  very likely that a randomly selected sample is close to zero and this
  observation is readily explained from the fact that $a_p(t)$ has the same
  fall-off rate as a Gaussian pulse, $\sim \exp(-t^2)$. Nevertheless, there are
  some samples with very high amplitude, although they are barely visible.
\end{itemize}
The discussion about a potential \vsi\ definition can start from the definitions
of the standardized moments of a probability
distribution.\footnote{\url{https://en.wikipedia.org/wiki/Standardized_moment}}
\begin{itemize}
\item The first standardized moment, $\tilde{\mu}_1$, is the first moment of the
  probability distribution function (PDF) around the mean, which is identically
  zero.
\item The second standardized moment, $\tilde{\mu}_2$, is one because it is the
  second moment about the mean (i.e., the variance) normalized to the variance,
  $\sigma^2$.
\item The third standardized moment is a measure of skewness and this quantity
  does not seem to be of use in this case.
\item The fourth standardized moment, defined according to $\tilde{\mu}_4 =
  \mu_4/\sigma^4$, is promising and is further described below.
\end{itemize}
The fourth standardized moment is known as the \emph{kurtosis} and is related to
the asymptotic fall-off rate of a PDF. The (univariate) normal distribution has
a kurtosis of 3 and it is common to use the \emph{excess kurtosis}, which is the
kurtosis minus 3. A distribution with negative excess kurtosis is called
\emph{platykurtic} and the most extreme example of such a PDF is the Bernoulli
distribution with $p = 1/2$. This is the discrete PDF for a random variable that
takes the values 0 and 1 with equal probability and has an excess kurtosis of
$-2$. A continuous uniform distribution, where the PDF is constant for a certain
interval and zero outside this interval, has an excess kurtosis of $-6/5$. The
numerically obtained kurtosis for the continuous signal is \num{-1.5}, which is
not surprising as its PDF can (in a vague sense) be said to be somewhere between
the two distributions discussed above.

A distribution with positive excess kurtosis, on the other hand, is called
\emph{leptokurtic}. The excess kurtosis is high when there are many samples that
occur far away from the mean value, when the distance is measured in standard
deviations of the PDF. One example of a leptokurtic distribution is the Laplace
distribution, which falls off as $\sim \exp(-|x|)$ as compared to $\sim
\exp(-x^2)$ for the normal distribution. The excess kurtosis for the Laplace
distribution is \num{3}, but this is very low compared to the value obtained for
the pulsed model signal, which is \num{35.6}. The reason for this is that the
standard deviation is small due to the accumulation of samples close to zero,
but there are some samples with high values and these increase the excess
kurtosis significantly.

It is noted that WGN will be classified in a reasonable way as the excess
kurtosis of squared WGN is about 12, i.e., in between the values for a
continuous and a pulsed signal. Furthermore, if $T_p$ is decreased, then the
excess kurtosis becomes comparable to the value for the continuous model signal.
For example, if the pulse width is set to $T_p = \SI{0.01}{s}$, which makes the
continuous and pulsed model signals quite similar in appearance, then the excess
kurtosis of $a_p^2$ becomes \num{-0.23}. If, on the other hand, $T_p$ is made
larger than \SI{0.1}{s}, then also the excess kurtosis is increased.

This suggests that the excess kurtosis could be used for defining the \vsi. A
drawback of this, however, is that although the excess kurtosis is a well-known
statistical concept that is used in many contexts, it may still be unknown to a
large fraction of engineers and technicians. It is also fair to say that even
with knowledge about the kurtosis, it is still somewhat hard to develop an
intuitive understanding for the \vsi\ when defined in this way. Furthermore, it
is not obvious how the corresponding \vsl\ could be defined if the \vsi\ is
based on kurtosis.

\subsection{Using Cumulative Energy---``Energy Steps''}
\begin{figure}[p]
  \begin{center}
    \includegraphics[width=0.75\linewidth]{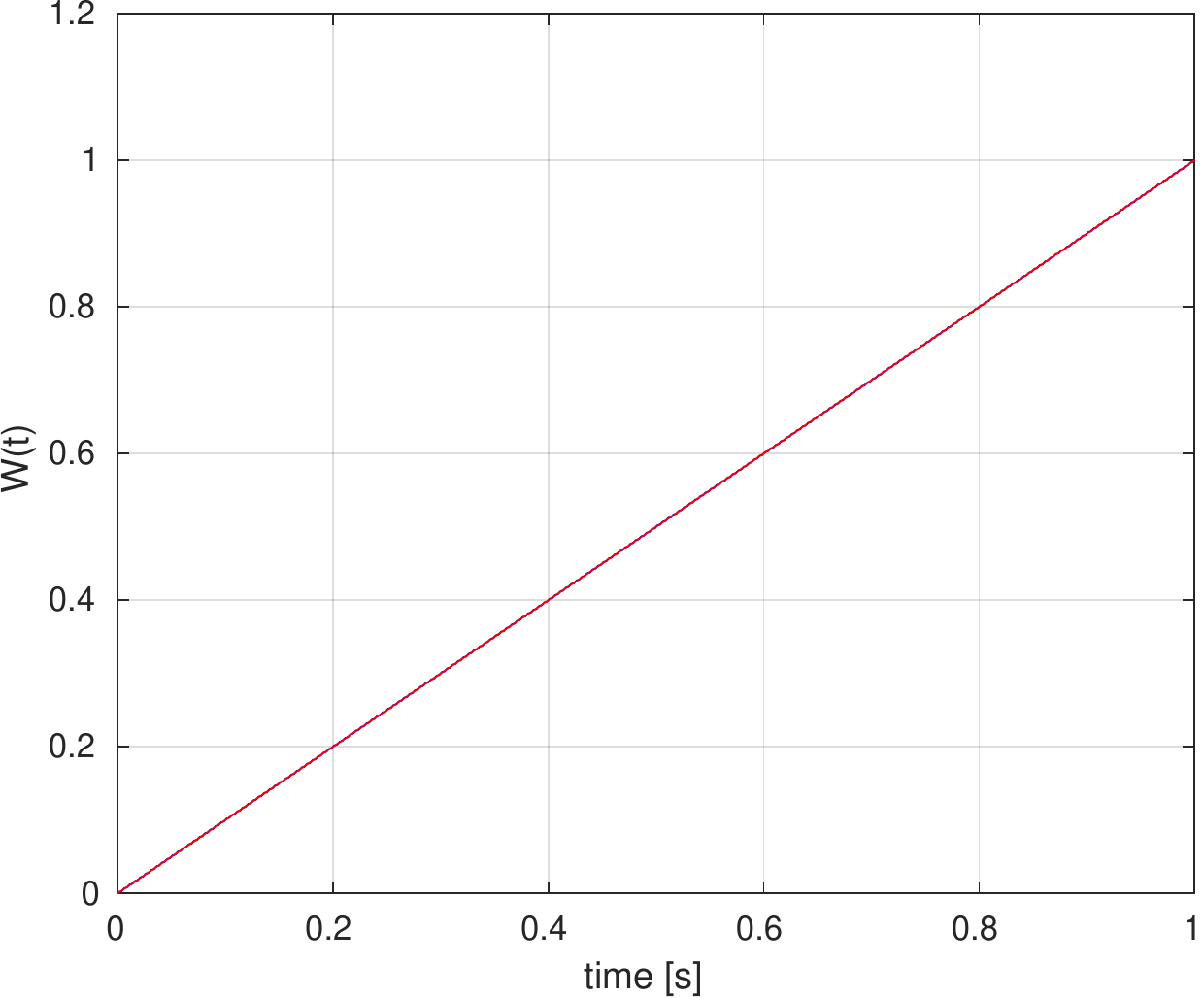}
    \caption{The $W(t)$ function for the continuous model signal.}
    \label{fig_W_t_continuous_signal}
  \end{center}
\end{figure}

\begin{figure}[p]
  \begin{center}
    \includegraphics[width=0.75\linewidth]{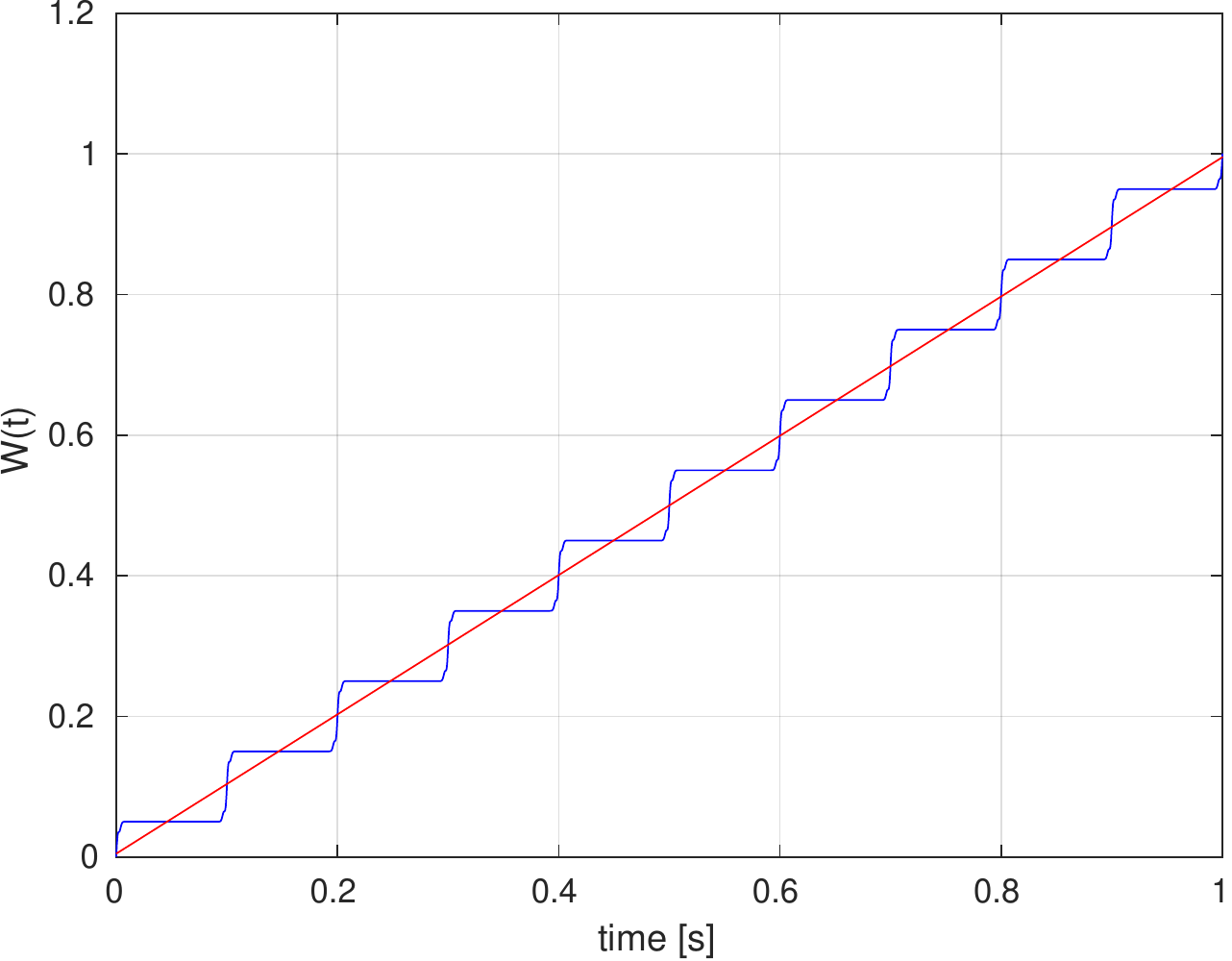}
    \caption{The $W(t)$ function for the pulsed model signal.}
    \label{fig_W_t_pulsed_signal}
  \end{center}
\end{figure}

Instead trying to build more directly on the argument from
Section~\ref{sec:Problem Formulation}, the \vsi\ should be high if few samples
in the signal carry a large part of the signal energy. This means that for a
signal that contains powerful pulses, the energy signal, defined
\begin{align}
  W(t) = \int_{-\infty}^{t} P(\tau) \, d\tau = \int_{-\infty}^{t} a^2(\tau) \, d\tau,
\end{align}
is expected to be shaped approximately like a staircase. To exemplify this,
$W(t)$ has been plotted for the continuous and the pulsed model signals in
Figs.~\ref{fig_W_t_continuous_signal} and \ref{fig_W_t_pulsed_signal}. It is
clear that in the case of the pulsed model signal, $W(t)$ is close to being
step-shaped.

To discuss this case in detail, an ideal step-shaped function is defined
according to
\begin{align}
  W(t) = k \, \Delta W, \quad |t - k \, \Delta t| < \Delta t/2, \quad \forall k,
\end{align}
where $\Delta W$ is the ``energy step'' and $\Delta t$ is the time duration of
each step. For an ideal continuous signal, the energy function would instead be
linear, $W(t) \propto t$. The best linear approximation to both types of curves
would be
\begin{align}
  \tilde{W}(t) = \frac{\Delta W}{\Delta t} t,
\end{align}
where $\Delta W/\Delta t$ in both cases can be obtained by fitting a straight
line. The error can then be defined as the difference between the energy signal
and its linear fit according to
\begin{align}
  \epsilon \equiv W(t) - \tilde{W}(t) = W(t) - \frac{\Delta W}{\Delta t} t
\end{align}
and the mean square error (MSE) can be analytically calculated for the
step-shaped function according to
\begin{align}
  \langle \epsilon^2 \rangle = \frac{\Delta W^2}{12}.
\end{align}
This implies that it is possible to find the effective energy step and for a
signal that is not step-shaped, this quantity is expected to be small.

While this may seem like a promising way to define the \vsi, the calculated
quantity, $\Delta W$, is not dimensionless, which was one of the criteria
demanded for the \vsi\ above. In this case, $\Delta W$ will change, e.g., if the
acceleration signal is rescaled, which is not surprising as $\Delta W$ is the
pulse energy (in arbitrary units). This was not what was intended to be
quantified. A possible measure for how localized the signal is could potentially
be found by relating the pulse energy to the signal mean energy, but this must then
be interpreted as \emph{mean energy per pulse period} and this, consequently,
raises the question what the pulse period is, which may not have an unambiguous
answer and for a non-pulsed signal (such as WGN) the concept
\emph{pulse period} may not even be meaningful. Thus, it is concluded that
while the concept of calculating the ``energy steps'' is interesting, it seems
difficult to base the \vsi\ definition on it.

\subsection{Using the Power Signal Distribution}
\label{sec:Using the Power Signal Distribution}
Instead trying to apply the argument from Section~\ref{sec:Problem Formulation}
to the power signal histograms discussed in Section~\ref{sec:Using the Excess
  Kurtosis}, there should, in a relative sense, be few samples with a large
value in the pulsed model signal. On the other hand, as seen in
Fig.~\ref{fig_hist_pulsed_signal}, there should be many samples with an
amplitude close to zero. Then, by selecting a threshold and counting the number
of samples below, $n_\text{low}$, and above, $n_\text{high}$, this threshold,
the \vsi\ could be defined
\begin{align}
  \text{\vsi} \equiv \frac{n_\text{low}}{n_\text{high}}.
\end{align}
If the threshold is set to \SI{50}{\percent} of the maximum $a^2$ value, it is
obtained that $\text{\vsi} = 1$ for a harmonic oscillation and $\text{\vsi}
\approx 47$ for the pulsed model signal. Thus, this parameter may seem
promising, but one problem is that it is very sensitive to outliers, which can
be exemplified by using WGN. It is then found that the \vsi\ is very high
($\gtrsim 1000$), although WGN is not localized in time. Even worse, the \vsi\
calculated for WGN is varying significantly between simulations and the result
may be exceptionally large when an unlikely high maximum value happens to have
been drawn. As a matter of fact, the normal distribution has a finite
probability for \emph{any} value of the amplitude and this implies that any
parameter based on the maximum signal value has undesirable statistical
properties. Thus, the requirement for a robust definition of the \vsi\ excludes
any use of the maximum of the power signal.

\subsection{Using Cumulative Energy from the Power Signal Distribution}
\label{sec:Using Cumulative Energy from the Power Signal Distribution}
\begin{figure}[p]
  \begin{center}
    \includegraphics[width=0.75\linewidth]{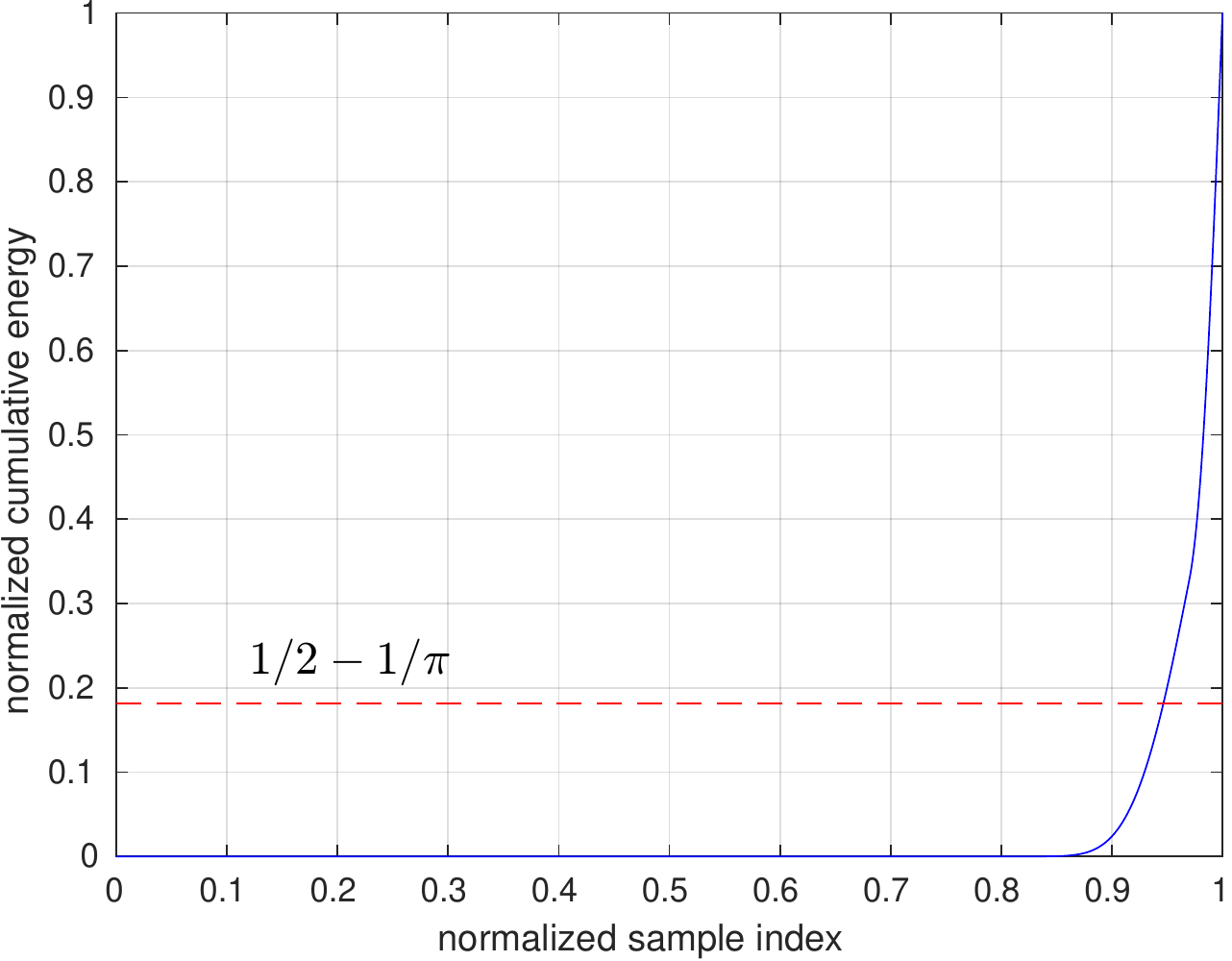}
    \caption{The cumulative sum of the sorted power of the pulsed model signal.}
    \label{fig_cumsum_sort_power_pulsed_signal}
  \end{center}
\end{figure}

To handle the problem with outliers encountered in Section~\ref{sec:Using the Power Signal Distribution}, the approach can be modified to be based on cumulative energy instead of power samples according to the following.
\begin{enumerate}
\item The power signal samples are calculated and sorted.
\item The sorted power signal is integrated into the cumulative energy using a
  cumulative sum, i.e., a sequence of partial sums.
\item Using a selected threshold, $W_\text{th}$, for the cumulative energy, the
  number of samples below and above the threshold are counted.
\item The \vsi\ is calculated according to
  \begin{align}
    \text{\vsi} \equiv \frac{n_\text{low}}{n_\text{high}}.
  \end{align}
\end{enumerate}
A strict definition of this approach is found in Appendix~\ref{Algorithm} and a
Matlab implementation is found in Appendix~\ref{Algorithm Matlab
  Implementation}. Figs.~\ref{fig_cumsum_sort_power_continuous_signal} and
\ref{fig_cumsum_sort_power_pulsed_signal} show the cumulative sums using
normalized x- and y-axes. It is clearly seen that for the pulsed model signal,
few samples contribute a large fraction of the total signal energy. One
advantage with this definition of the \vsi\ is that it is in direct
correspondence with the criterion described in Section~\ref{sec:Problem
  Formulation}.\footnote{The essential difference from the ``energy step''
  method is that the signal is sorted before it is integrated.} Outliers should
not affect this quantity as a negligible part of the total energy should be
contributed by these.

\begin{figure}[p]
  \begin{center}
    \includegraphics[width=0.75\linewidth]{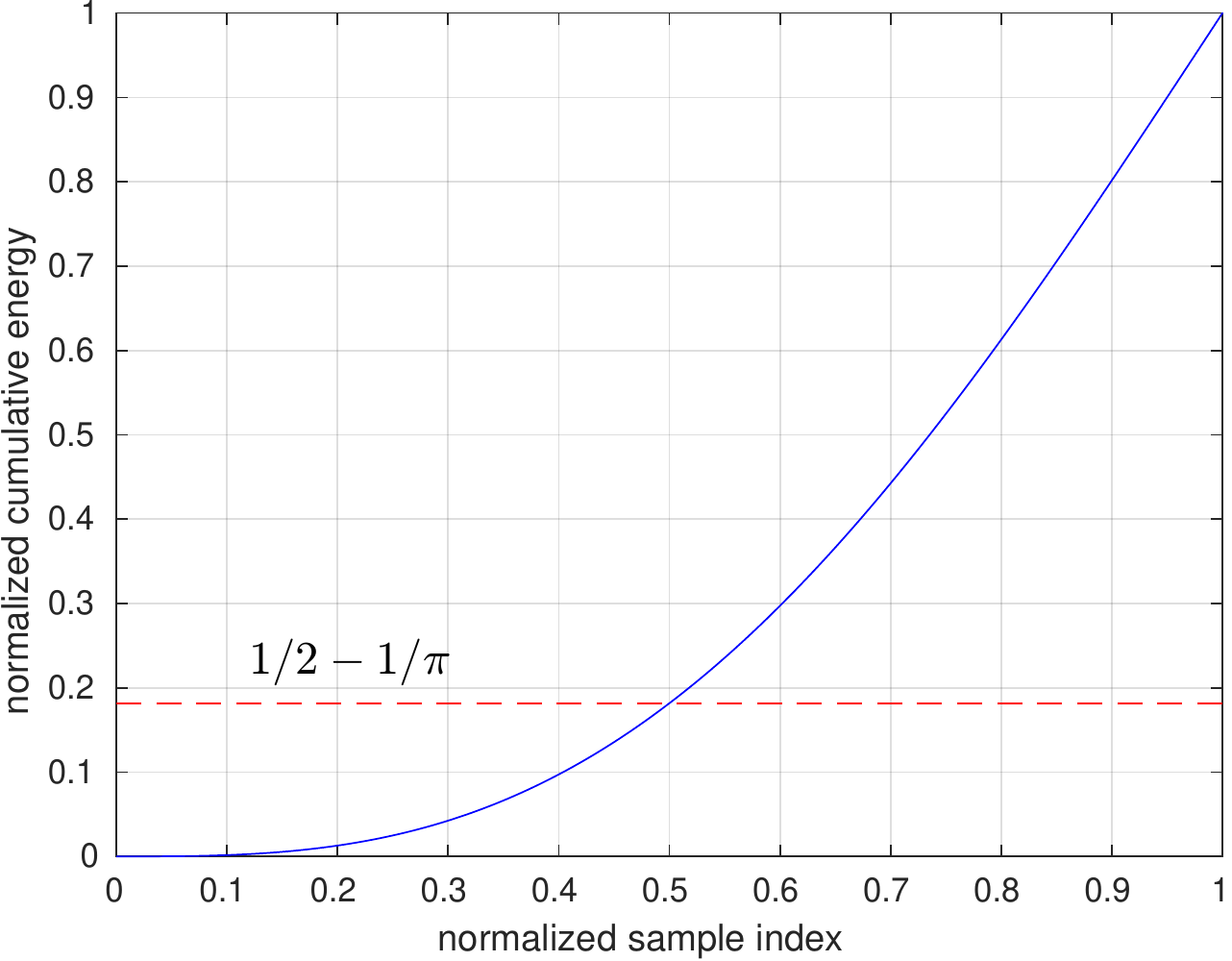}
    \caption{The cumulative sum of the sorted power of the continuous model
      signal.}
    \label{fig_cumsum_sort_power_continuous_signal}
  \end{center}
\end{figure}

It remains to select the threshold for the cumulative energy. One obvious
suggestion is to set the level at \SI{50}{\percent} of the total energy,
$W_\text{tot}$. This should work well but one alternative is to select the
threshold such that the \vsi\ for a harmonic oscillation is one. The
calculations for this are found in Appendix~\ref{sec:Sample Distribution for a
  Harmonic Signal} and \ref{Cumulative Energy from the Power Signal
  Distribution} and the result is that by selecting the threshold value
according to
\begin{align}
  \frac{W_\text{th}}{W_\text{tot}} = \frac{1}{2} - \frac{1}{\pi} \approx 0.18,
\end{align}
then $\text{\vsi} = 1$ for a harmonic oscillation. For the pulsed model signal,
it is obtained that $\text{\vsi} \approx 17.7$ and, crucially, for WGN it is
obtained that $\text{\vsi} \approx 2.0$. Thus, compared to a harmonic
oscillation WGN gives a somewhat higher but comparable value and the value for
the pulsed model signal is much larger.

It should be noted that the cumulative energy plots makes it possible to discuss
the signal in other intuitive ways. For example, in
Fig.~\ref{fig_cumsum_sort_power_continuous_signal} it is seen that the curve
passes very close to the point $(0.9, 0.8)$. The interpretation of this is that
10\,\% of the signal samples carry 20\,\% of the energy. In
Fig.~\ref{fig_cumsum_sort_power_pulsed_signal}, on the other hand, the curve
reaches the 80\,\% energy level approximately at the $x$-coordinate 0.994, i.e,
20\,\% of the signal energy are carried by 0.6\,\% of the samples.

% It remains to make an investigation of the \vsi\ values that result from
% experimental signals, but preliminary results have been promising. For example,
% a sander was found to have a \vsi\ in between those for a harmonic oscillation and
% WGN. Higher values were obtained for a reciprocating saw, \vsi\ = 3.7, and an
% impact wrench, \vsi\ = 10.0. These results, the discussion above, and the fact
% that all the design criteria above seem to be fulfilled makes this candidate
% definition of the \vsi\ interesting.

Having found a candidate \vsi\ definition, a method for calculating \vsl\ is
needed. The intention for this parameter is \emph{not} to produce the peak
values for the shock events in the vibration signal. Such a parameter would be
sensitive to outliers and even in the absence of outliers, the value would be
unreasonably high if there is a single event in the signal with very high
amplitude. Instead, a \emph{characteristic} value for the shocks is sought and
this can be obtained from the \vsi\ procedure described above.

One method is simply to define the \vsl\ as the square root of the sorted power
signal at the threshold energy level. Using the same definition of $W_\text{th}$
as above the \vsl\ value for the continuous signal is \SI{1}{m/s^2}, i.e., the
RMS value. The \vsl\ value for the pulsed signal is \SI{2.4}{m/s^2}. The \vsl\
will for both signals be more comparable to the peak values if the $W_\text{th}$
is selected at the \SI{50}{\percent} level. The ratio of the peak amplitudes is
\num{4.1}. For WGN, the \vsl\ is very similar to the RMS value.

\subsection{Using a Weighted Mean Square Value}
\label{sec:Using Weighted Mean Square Values}
A different approach to using the criterion from Section~\ref{sec:Problem
  Formulation} is to ask if the parts of the signal that carry a large part of
the signal energy can be identified and isolated in some way. The aim of this
section is to use weighted mean squares to perform this. When doing this, it is
more natural to first seek a definition of the \vsl\ and then the \vsi\ can be
based on the \vsl.

The averaging method used in ISO~5349 does not yield a result that is
representative for the shock parts in a signal with high \vsi. A different
method is therefore needed but as weighted mean squares can be viewed as a
generalization of the method used in the standard, a brief overview of the
standardized method will first be given.

\subsubsection{Quantification Method in ISO~5349}
It is here convenient to think about an idealized situation where the
acceleration is measured with a perfect instrument without any bandwidth
limitation. The quantification of a vibration measurement according to ISO~5349
is then performed in two steps. First, the signal is filtered. Then, the ISO
value, $a_\text{hv}$, is calculated using the RMS values for the three
orthogonal axes according to ISO~5349-1~\cite[Section~4.5]{iso5349-1},
\begin{align}
  a_\text{hv} = \sqrt{a^2_{\text{hw}x} + a^2_{\text{hw}y} + a^2_{\text{hw}z}}.
\end{align}
The filtering operation can be written
\begin{align}
  \mathcal{F}^{-1} \left[ \mathcal{F} \left[ \cdot \right] H(f) \right],
\end{align}
where the Fourier transformation is denoted by $\mathcal{F}[\cdot]$ and $H(f)$
is the transfer function defined in ISO~5349-1~\cite[Annex A]{iso5349-1}. In order
to quantify the impact of ultravibrations this filter will need to be modified
as the high frequency content otherwise is lost, but the discussion about how to
select this is outside the scope of this report.

\subsubsection{Weighted Vibration Signal Averaging}
From the definition of the root mean square concept, the mean square (MS) of a
vibration signal can be written
\begin{align}
  \text{MS} = \frac{\int a^2(t) \, dt}{\int dt} = \frac{1}{T} \int a^2(t) \, dt,
\end{align}
where the integration is performed over a representative measurement of the
signal of duration $T$. If the signal is pulsed, then the RMS value has no
direct connection to the signal values occurring within the pulses. In fact, by
moving the pulses further and further apart, the RMS value would approach zero.
The question is then how a value that represents the pulses can be calculated.
One approach is to modify the RMS value into a weighted value, where the
weighting puts special emphasis on the pulses. In general, a weighted time
average of a signal $u(t)$ is calculated according to
\begin{align}
  \frac{\int u(t) w(t) \, dt}{\int w(t) \, dt},
\end{align}
where $w(t)$ is a suitably chosen weight function. It is seen that the mean
square calculation of ISO~5349 uses $w(t) = 1$ and $u(t) = a^2(t)$, which
corresponds to weighting all parts of the signal equally. The weighted mean
square (WMS) concept is introduced in a general way in Appendix~\ref{Weighted
  Averaging}, where also a number of examples of well-known applications are
given.

In order to quantify the shocks, the power signal should be averaged, i.e.,
$u(t) = P(t) = a^2(t)$. Emphasis should be put on the parts of the signal where
the power is high and this suggests that the weight function should be a
monotonous function that is increasing with the power level. There is an
infinite number of such functions and it is not obvious how to make a selection.
The guiding principle here is that the weighting function should be as simple as
possible but still produce reasonable numerical values for the pulse amplitude.

The simplest possible function is to select $w(t)$ to a constant value, but in
order to average the high-power parts of the signal, a piecewise constant
function is needed. By selecting $w(t)$ to be one for the high-power parts and
zero elsewhere, the weighting function becomes a windowing function that removes
the low-power parts completely. One obvious problem with this is how the
threshold value for the power signal should be selected. In fact, if such a
value can be selected, the definition of the \vsl\ could very well be based on
it, leading to a circular argument.

Using monomials is another very simple choice of weighting functions, giving 
\begin{align}
  w(t) = P^K(t) = a^{2 K},
\end{align}
where $K$ is not necessarily an integer. The MS definition in the standard
corresponds to selecting $K = 0$, but selecting a larger value will make the
weight function narrower, thereby selecting a smaller, more intense part of the
shock. Increasing $K$ to a large value would lead to extreme weighting that
will, in effect, result in the maximum of the amplitude. As this is undesirable,
it seems reasonable to limit $K$ to a small value. For example, selecting $K =
2$, the WMS becomes
\begin{align}
  \text{WMS} = {\frac{\int a^2(t) P^2(t)\, dt}{\int P^2(t) \, dt}}
  = {\frac{\int a^6(t) \, dt}{\int a^4(t) \, dt}}
\end{align}
% A simple choice is to select $w(t) = P(t) =
% a^2(t)$, leading to a weighted mean square (WMS) according to
% \begin{align}
%   \text{WMS} = \frac{\int a^2(t) P(t)\, dt}{\int P(t) \, dt} = \frac{\int a^4(t) \, dt}{\int a^2(t) \, dt}
% \end{align}
and the \vsl\ is then defined as a root weighted mean square according to
\begin{align}
  \text{\vsl} = \text{RWMS} = \sqrt{\text{WMS}}.
\end{align}
A straightforward way to define the \vsi\ is then
\begin{align}
  \text{\vsi} = \text{RWMS}/\text{RMS},
\end{align}
where the RMS value is obtained by using $w(t) = 1$.

Selecting $K = 2$, the $\text{\vsi} \approx 1.29$ for a harmonic oscillation.
For the pulsed model signal, it is obtained that $\text{\vsi} \approx 5.1$ and
for WGN it is obtained that $\text{\vsi} \approx 2.2$. Compared to the
cumulative energy method, the separation of the \vsi\ values is smaller, but in
a relative sense, both methods give fully reasonable results. The \vsl\ value
for the continuous signal is \SI{1.3}{m/s^2}. The \vsl\ value for the pulsed
signal is \SI{5.1}{m/s^2}, i.e., \num{3.9} times higher. The ratio of the peak
amplitudes is \num{4.1}. For WGN, the \vsl\ is \SI{2.2}{m/s^2}. Also these
values follow what is intuitively expected.

\section{Conclusion}
The aim of this document has been to suggest a definition for the
\emph{\vsifull} and the \emph{\vslfull} parameters. These quantify to what
extent a given vibration signal is to be considered to contain
shocks/impacts/transients and what a characteristic vibration level of these
events within the signal is, respectively. Among the investigated candidate
definitions, the ones based on the excess kurtosis, the cumulative energy from
the power signal distribution, and the weighted mean squares, seem to fulfill
the criteria stated in Section~\ref{sec:Selection Criteria}. The candidate
definition based on excess kurtosis seems less intuitive than the alternatives
and also seems to make a definition of the \vsl\ difficult. The candidate
definitions based on cumulative energy and weighted mean squares are both
well worth evaluating on experimental signals.

%%%%%%%%%%%%%%%%%%%%%%%%%%%%%%%%%%%%%%%%%%%%%%%%%%%%%%%%%%%%%%%%%%%%%% 
% Reference formatting depends on publisher and field. The IEEE
% transactions style is a good choice for text published, e.g., at
% Chalmers or KTH.
\bibliographystyle{IEEEtran}
\bibliography{vsi}

%%%%%%%%%%%%%%%%%%%%%%%%%%%%%%%%%%%%%%%%%%%%%%%%%%%%%%%%%%%%%%%%%%%%%% 
% Appendices
% \newpage
\appendix
\section{Dissipated Energy from a Harmonic Oscillator}
\label{sec:Dissipated Energy from a Harmonic Oscillator}
The model equation for a driven harmonic oscillator is
\begin{align}
  m \ddot{x} + c \dot{x} + k x = F(t),
\end{align}
where $m$ is the mass, $c$ is the damping, $k$ is the spring constant, and $F$ is the drive force. The instantaneous power delivered by the external force is $F v$, i.e.,
\begin{align}
  m \ddot{x} \dot{x} + c \dot{x}^2 + k x \dot{x} = F(t) \dot{x} = P(t),
\end{align}
which is rewritten to
\begin{align}
  \frac{m}{2} \frac{d \dot{x}^2}{dt} + c \dot{x}^2 + \frac{k}{2} \frac{d x^2}{dt} = P(t).
\end{align}
Integrating $\int_{-\infty}^{t} \! \! \cdot \, dt$ gives
\begin{align}
  \frac{m}{2} [\dot{x}^2]_{-\infty}^t + c \int_{-\infty}^{t} \dot{x}^2 \, dt + \frac{k}{2} [x^2]_{-\infty}^{t} = \int_{-\infty}^{t} P(t') \, dt' = W_\text{tot}(t)
\end{align}
and assuming that $x \to 0$ and $\dot{x} \to 0$ as $t \to -\infty$, it is obtained that
\begin{align}
  W_k(t) + c \int_{-\infty}^{t} \dot{x}^2 \, dt + W_p(t) = W_\text{tot}(t).
\end{align}
Systems of the kind considered here operate a long time and the kinetic energy, $W_k$, and the potential energy, $W_p$, are bounded and oscillatory. This implies that after the initial transient, all energy delivered by the driving force will be dissipated by the damping term. It is true that the energy dissipation mainly occurs in the material being machined and in the vibrating equipment itself, but it is reasonable to assume that the energy dissipated in the operator is also proportional to the squared velocity, i.e.,
\begin{align}
  P_\text{hand} \propto \dot{x}^2
\end{align}
with the corresponding dissipated energy
\begin{align}
  W_\text{hand} \propto \int \dot{x}^2 dt.
\end{align}

\section{Model Signal Calculations}
\label{sec:Model Signal Calculations}
The continuous signal
\begin{align}
  x_c(t) &= A_c \cos(2 \pi f_c t + \phi_c), \nonumber \\
  v_c(t) &= -A_c (2 \pi f_c) \sin(2 \pi f_c t + \phi_c), \nonumber \\
  a_c(t) &= -A_c (2 \pi f_c)^2 \cos(2 \pi f_c t + \phi_c).
\end{align}
The pulsed signal
\begin{align}
  x_p(t) &= \sum_{k=-\infty}^{\infty} A_p \exp \left[ - \frac{(t - k T_p)^2}{2 t_p^2} \right],\nonumber \\
  v_p(t) &= -\sum_{k=-\infty}^{\infty} A_p \frac{(t - k T_p)}{t_p^2} \exp \left[ - \frac{(t - k T_p)^2}{2 t_p^2} \right] \nonumber \\
  a_p(t) &= \sum_{k=-\infty}^{\infty} A_p \frac{(t - k T_p + t_p)(t - k T_p - t_p)}{t_p^4} \exp \left[ - \frac{(t - k T_p)^2}{2 t_p^2} \right].
\end{align}
The pulse width $t_p$ is selected such that the Fourier transform of $a_p$ has its maximum at $f_c$, which gives
\begin{align}
  t_p = \frac{1}{\sqrt{2} \, \pi f_c}.
\end{align}
The RMS value for the continuous signal is
\begin{align}
  a_{c, \text{RMS}}^2 = \frac{1}{T_c} \int_{-T_c/2}^{T_c/2} a_c^2(t) \, dt = 8 \pi^4 f_c^4 A_c^2.
\end{align}
Assuming that $t_p \ll T_p$, the corresponding expression for the pulsed signal is
\begin{align}
  a_{p, \text{RMS}}^2 = \frac{1}{T_p} \int_{-\infty}^{\infty} a_p^2(t) \, dt = \frac{3 \sqrt{\pi} \, A_p^2}{4 T_p t_p^3}.
\end{align}
Setting these equal, it is found that
\begin{align}
  A_p = 4 \sqrt{\frac{f_c T_p}{3} \sqrt{\frac{\pi}{2}}} \, A_c.
\end{align}

\section{Algorithm based on Cumulative Energy}
\subsection{Algorithm Definition}
\label{Algorithm}
The purpose of this section is to give an unambiguous definition of the algorithm introduced in Section~\ref{sec:Using Cumulative Energy from the Power Signal Distribution}. Assume that the measurement of the acceleration has resulted in the samples $a_n$, $n = 1,2,\ldots,N$. The power signal samples are then $P_n = a_n^2$ and the sorting results in a reordered set of power signal samples denoted $\tilde{P}_n$, $n = 1,2,\ldots,N$. The cumulative energy is calculated in terms of the $N$ partial sums
\begin{align}
  W_M \equiv \Delta t \sum_{n = 1}^{M} \tilde{P}_n, \quad M = 1,2,\ldots,N,
\end{align}
where $\Delta t$ is the sampling interval. The largest value for $M$ for which $W_M < W_\text{th}$ is found and the \vsi\ is calculated according to
\begin{align}
  \text{\vsi} = \frac{M}{N - M}.
\end{align}
This can be expressed simpler using the total energy, which is the $N$th partial sum
\begin{align}
  W_\text{tot} = \Delta t \sum_{n = 1}^{N} \tilde{P}_n.
\end{align}
Introducing normalized partial sums
\begin{align}
  \widehat{W}_M \equiv \frac{W_M}{W_\text{tot}} = \frac{\sum_{n = 1}^{M} \tilde{P}_n}{\sum_{n = 1}^{N} \tilde{P}_n}, \quad M = 1,2,\ldots,N,
\end{align}
the largest value for $M$ for which $\widehat{W}_M < W_\text{th}/W_\text{tot}$
should instead be found. It should be noted that $\widehat{W}_M$ is the
normalized cumulative energy plotted in
Figs.~\ref{fig_cumsum_sort_power_continuous_signal} and
\ref{fig_cumsum_sort_power_pulsed_signal} as a function of the normalized sample
index vector, i.e., the vector $1/N,2/N,\ldots,1.$

\subsection{Matlab Implementation}
\label{Algorithm Matlab Implementation}
\begin{minted}{octave}
function [vsi, vsl, n_norm, W_norm] = algorithmCumulEnergy(a)
    n = 1:length(a); % Sample index vector
    n_norm = n/length(n);
    P = a.^2; % Power signal
    P_sort = sort(P);
    W = cumsum(P_sort); % Cumulative energy
    W_norm = W/W(end);
    % W_th = 0.5;
    W_th = 1/2 - 1/pi;
    n_low  = sum(W_norm < W_th); % Number of samples below threshold
    n_high = sum(W_norm > W_th); % Number of samples above threshold
    vsi = n_low/n_high;

    idx = find(W_norm > W_th); % Samples above threshold
    % VSL from the smallest power signal sample above threshold
    vsl = sqrt(P_sort(idx(1)));
end
\end{minted}

\section{Harmonic Signal}
\subsection{Sample Distribution}
\label{sec:Sample Distribution for a Harmonic Signal}
If the signal
\begin{align}
  y = \sin{t}
\end{align}
is sampled at a random time, the sample will be drawn from a certain distribution. To find this distribution, the function is first restricted to the interval $t \in [-\pi/2, \pi/2]$, where the function is monotonous. The sample distribution, $f(y)$, by definition fulfills
\begin{align}
  \text{Pr} [y_1 \le y \le y_2] = \int_{y_1}^{y_2} f(y) \, dy.
\end{align}
The sampling time is assumed to be drawn from a uniform random distribution, implying that the probability that the sample is drawn in a small interval $dt$ is $dt/\pi$. The $y$-value will then be in the interval $[y, y + dy]$, for which the probability by definition is $f(y) \, dy$. Setting these two probabilities equal gives
\begin{align}
  f(y) = \frac{1}{\pi} \frac{dt}{dy} = \frac{1}{\pi} \frac{d}{dy} \arcsin (y) = \frac{1}{\pi} \frac{1}{\sqrt{1 - y^2}}.
\end{align}

\subsection{Cumulative Energy}
\label{Cumulative Energy from the Power Signal Distribution}
For a power signal that is a harmonic oscillation, it is possible to calculate the cumulative energy analytically. First, it is noted that the energy signal, which is the square of the power signal is also harmonic, although it is oscillating from zero to a maximum value. Thus, the PDF for the power signal has the shape derived in Appendix~\ref{sec:Sample Distribution for a Harmonic Signal}, although the limits for the distribution should be shifted from $[-1, 1]$ to $[0, P_\text{max}]$. Instead of shifting the distribution, the power corresponding to the original PDF is introduced as $(P + 1)$, i.e., a linearly increasing function starting at zero. The energy is then obtained by integration according to
\begin{align}
  W &= \int (P + 1) f(P) \, dP \nonumber \\
    &= \int (P + 1) \frac{1}{\pi} \frac{1}{\sqrt{1 - P^2}} \, dP \nonumber \\
    &= \frac{1}{\pi} \left( \int \frac{1}{\sqrt{1 - P^2}} \, dP + \int \frac{P}{\sqrt{1 - P^2}} \, dP \right) \nonumber \\
    &= \frac{1}{\pi} \left(\arcsin(P) - \sqrt{1 - P^2} \right).
\end{align}
Integrating over the entire interval, it is obtained that
\begin{align}
  W_\text{tot} &= \int_{-1}^{1} (P + 1) f(P) \, dP = 1.
\end{align}
Instead, integrating over half of the samples, the result is
\begin{align}
  W &= \int_{-1}^{0} (P + 1) f(P) \, dP \nonumber \\
    &= \frac{1}{\pi} \left[ \left(\arcsin(0) - \sqrt{1 - 0} \right) - \left(\arcsin(-1) - \sqrt{1 - 1} \right) \right] \nonumber \\
    &= \frac{1}{\pi} \left( \frac{\pi}{2} - 1 \right) = \frac{1}{2} - \frac{1}{\pi} \approx 0.18.
\end{align}
This means that if the signal is harmonic and the threshold for the cumulative
energy is set to $W_\text{th}/W_\text{tot} = 1/2 - 1/\pi$, then there will be an equal number of samples above and below the threshold.

\section{Weighted Averaging}
\label{Weighted Averaging}
The aim of this section is to give an introduction to weighted averaging and
provide an intuitive explanation of the expression
\begin{align}
  \label{wms}
  \frac{\int u(t) w(t) \, dt}{\int w(t) \, dt},
\end{align}
which was used as a starting point for the discussion about the WMS value
above.

The arithmetic mean value of a set of samples $x_i$, $i = 1, 2,\ldots,N$ is
defined
\begin{align}
  \bar{x} = \frac{1}{N} \sum_{i = 1}^N x_i
  = \frac{\sum_{i = 1}^N x_i \times 1}{\sum_{i = 1}^N 1},
\end{align}
where the second expression is written on a form that is more similar to the
weighted expression in (\ref{wms}). In this case, it is $x_i$ that is being
averaged and all samples are weighted equally. If the values among the samples
$x_i$ occur $p_i$ times (instead of a single time), then the summation can be
performed over the (different) values instead of over the samples. One example
is a die which can land on only one of the six sides. The averaging operation
is then modified to a weighted form according to
\begin{align}
  \bar{x} = \frac{\sum_{j = 1}^M x_j p_j}{\sum_{j = 1}^M p_j},
\end{align}
where $M$ is the number of different values. This is a discrete version of the
WMS expression in (\ref{wms}). Continuous WMS expressions, which are exactly on
the form of (\ref{wms}) are also common in statistics and physics. One example
is the expectation of a random variable $X$, which is defined
\begin{align}
  \mathbb{E} [X] = \frac{\int_{-\infty}^\infty x f(x) \, dx}{\int_{-\infty}^\infty f(x) \, dx},
\end{align}
where $f(x)$ is the probability density function (PDF) for $X$. Usually, the
denominator is not included in the definition, but this is because the PDF is
assumed to be normalized, i.e., the denominator is one. A very similar
expression is the definition for the center of mass, which is
\begin{align}
  \mathbf{R} = \frac{\iiint \mathbf{r} \rho \, dV}{\iiint \rho \, dV}
  = \frac{1}{M} \iiint \mathbf{r} \rho \, dV,
\end{align}
where $\rho$ is the density and the integration is over a continuous body. An
expression exactly on the form of (\ref{wms}) is obtained in the one-dimensional version
\begin{align}
  {R_x} = \frac{\int {x} \rho_x \, dx}{\int \rho_x \, dx},
\end{align}
where $\rho_x$ is the mass per length unit, not per volume unit. The
interpretation of this is that the center of mass is a weighted average where
the density acts as weighting function.

Similar to the center of mass definition, it is suggested in this report to use
a WMS value, where the weighting function emphasizes the parts with high power
level, to quantify the vibration level within the shocks. However, to further
draw parallels to the definition of the WMS expression above, it is actually
possible to also view the traditional mean square value (used to calculate the
RMS value) as a WMS value. To see this, the mean square value of $a^2$ is
written
\begin{align}
  \text{MS} = \frac{1}{T} \int_0^T a^2(t) \, dt =
  \frac{\int_0^T a^2(t) \, dt}{\int_0^T dt} = 
  \frac{\int_{-\infty}^\infty a^2(t) w(t) \, dt}{\int_{-\infty}^\infty w(t) \, dt}. 
\end{align}
In order to obtain equality in the last step, the weighting function is chosen
to be one in the interval $t \in [0, T]$ and zero otherwise. In this case, it
would be better to call $w(t)$ a \emph{windowing function}, but from the
expression it is clear that it is a WMS on the form above.

Many more examples of WMS expressions are found in the scientific literature,
but those given here should be a sufficient introduction to the calculation of
the shock amplitude.

\section{Algorithm based on Weighted Mean Squares}
\subsection{Algorithm Definition}
\label{Algorithm wms}
The purpose of this section is to give an unambiguous definition of the
algorithm introduced in Section~\ref{sec:Using Weighted Mean Square Values}.
Assume that the measurement of the acceleration has resulted in the samples
$a_n$, $n = 1,2,\ldots,N$. The power signal samples are then $P_n = a_n^2$ and
and the WMS is defined
\begin{align}
  \text{WMS}(K) = \frac{\sum_{n = 1}^N P_n P_n^K}{\sum_{n = 1}^N P_n^K},
\end{align}
where $K$ is the exponent in the weighting monomial function. The MS value is
obtained from
\begin{align}
  \text{MS} = \text{WMS}(0)
\end{align}
and the \vsl\ is defined
\begin{align}
  \text{\vsl} = \text{RWMS} = \sqrt{\text{WMS}}.
\end{align}
The \vsi\ is defined
\begin{align}
  \text{\vsi} = \frac{\text{RWMS}}{\text{RMS}} = \frac{\sqrt{\text{WMS}}}{\sqrt{\text{MS}}}.
\end{align}

\subsection{Matlab Implementation}
\label{Algorithm Matlab Implementation wms}
\begin{minted}{octave}
function [vsi, vsl] = algorithmWeightedMeanSquare(a)
    P = a.^2; % Power signal
    K = 2; % Monomial exponent for RWMS
    RWMS = sqrt(sum(P .* P.^K)/sum(P.^K));
    K = 0; % Monomial exponent for RMS
    RMS = sqrt(sum(P .* P.^K)/sum(P.^K));
    vsl = RWMS;
    vsi = RWMS/RMS;
end
\end{minted}

\end{document}